\documentclass[ reprint, amssymb, amsmath, aip,cha,floatfix]{revtex4-1}
\usepackage{bm}%
\usepackage{graphicx,color}
\usepackage{hyperref}%
\usepackage{enumerate}
\expandafter\ifx\csname package@font\endcsname\relax\else
 \expandafter\expandafter
 \expandafter\usepackage
 \expandafter\expandafter
 \expandafter{\csname package@font\endcsname}%
\fi
\begin{document}

\title{Reaching to a small target: entropic barriers and rates of specific binding of polymer to substrate}

\author{Samuel Bell and Eugene M. Terentjev}
\affiliation{Cavendish Laboratory, University of Cambridge,  J.J. Thomson Avenue, Cambridge, CB3 0HE, U.K.}
\begin{abstract}
This paper considers a broadly biologically relevant question of a chain (such as a protein) binding to a sequence of receptors with matching multiple ligands distributed along the chain. This binding is critical in cell adhesion events, and in protein self-assembly.
       Using a mean field approximation of polymer dynamics, we first calculate the characteristic binding time for a tethered ligand reaching for a specific binding site on the surface. This time is determined by two separate entropic effects: an entropic barrier for the chain to be stretched sufficiently to reach the distant target, and a restriction on chain conformations near the surface. We then derive the characteristic time for a sequence of single binding events, and find that it is determined by the `zipper effect', optimizing the sequence of single and multiple binding steps.
\end{abstract}

\maketitle

Ordered self-assembly requires the ability to organize and bind many molecules into a coherent structure. In biology, most self-assembling structures rely on specific interactions, matching ligands, and distinct binding sites. The kinetics of self-assembly is a broad and rich topic, which offers a fundamental understanding of processes being used in the construction of structured and functional aggregates.

One such process is the binding of a polymer chain to a sequence of receptor sites on a substrate. In considering adhesion of cells to each other, Jeppesen et al.~\cite{Jeppesen2001,Wong1997} had this problem for one specific binding site, where ligands tethered to the cell surface by flexible chains could also associate with the matching receptor on the adjacent cell. They found a dependence on the configuration of the polymer tether: in particular, how often the chains entered extended configurations to reach the distant receptors. Their treatment did not extend to an analytical expression of the binding rate, or multiple binding sites. Theoretically solving this problem is one of our main tasks here. 

\begin{figure}
	\centering
	\includegraphics[width=0.35\textwidth]{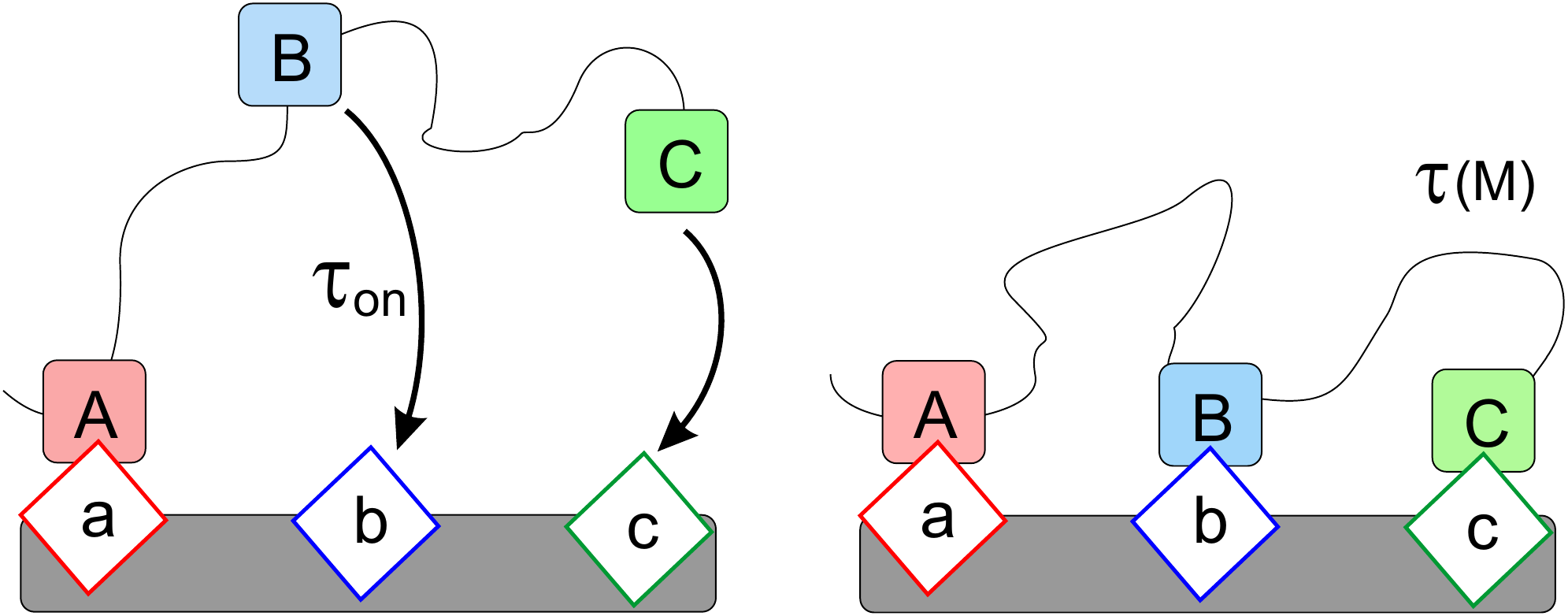}
	\caption{A scheme of ordered self-assembly of a polymer chain binding to a specific sequence of receptors on a substrate. Most of this paper is devoted to evaluation of the single binding time $\tau_\mathrm{on}$, which is a function of chain length A-B and the receptor separation a-b. We also examine how a total time $\tau(M)$ of assembly on a sequence of $M$ ordered binding sites emerges from the basic single-receptor problem. }
	\label{fig:binding}
\end{figure}

Another example of `reaching to a small target' is the polymer chain that could bind at multiple sites to the same surface, as in Fig. \ref{fig:binding}. Given a set of binding residues on the polymer chain, and a specifically placed set of receptors on the surface, what is the forward reaction constant of the full binding process? As a step towards the full problem, here we first consider the polymer after an initial binding event to the surface. This graft to the surface is persistent; we regard the tether as fixed. The remaining free chain has a second binding site, which undergoes thermal motion. It is able to bind to a second receptor also on the surface, a fixed distance $a$ away from the tethered origin. We calculate the mean time it takes the chain to `find' the target receptor, and discover that it is determined by an activation law where the effective potential barrier is purely entropic, $-T\Delta S$: as such, the explicit temperature disappears from the `activation law' and the mean binding time is proportional to $\exp[a^2/R_g^2]$, with $R_g$ the radius of gyration of the tethered chain. 

The search for a small target has already been considered in the context of polymer looping~\cite{FixmanI,FixmanII,Szabo1980,Pastor1996,Amitai2016}, where the time for two separate monomers on a polymer chain to meet was calculated. Such loops are observed experimentally in chromatin~\cite{chromatin1,chromatin2}. In fact, our calculation is based on the ideas of Szabo et al. \cite{Szabo1980}, although in their problem of forming a loop the distance to the binding site $a=0$ and accordingly no activation exponential has been observed. 
These types of problem are similar to the `narrow escape problem'~\cite{Bressloff2013,Holcman2014,Singer2006,Schuss2007}. Here a Brownian particle is confined to a domain whose boundary is entirely reflecting, apart from a small absorbing patch. The `narrow escape time' is the mean first time the particle reaches the absorbing patch and escapes the volume it was diffusing in.

When examining molecules confined to particular regions of space, we must consider carefully how this containment shows up in a mathematical framework. In the narrow escape problem for diffusing particles, the boundary is considered reflecting, with zero particle flux through the surface -- but in polymer physics the `hard' wall has a different effect due to the chain connectivity: the wall imposes an effective repulsive barrier making the monomer density zero on the wall. We discuss the ramifications of these boundary conditions by considering both a tethered polymer chain, comparing it with related problems in the literature.

The chain has a `hard' constraint of the wall to which it is grafted to, but its free end (with the binding ligand) also has a soft constraint on how far it can extend from the grafted origin. If the chain's end-to-end distance increases, there will be a resulting reduction in entropy~\cite{Guth1934}. Associated entropic barriers for activated processes have been investigated in polymer dynamics \cite{Muthukumar1988}, and in colloid glassy dynamics \cite{Schweizer2003}.
Entropic barriers have an important role to play in cell biology. They show up in polymer translocation through a membrane pore~\cite{Sung1996,Muthukumar1999}, as well as the mean looping time of a polymer chain \cite{Szabo1980}. They also play a role in the protein aggregation into amyloids \cite{Knowles2007}, and in more general protein folding funnel problems \cite{Szabo2000}. 

We also want to use these concepts to better understand the kinetics of adsorption of polymer chains to substrates. Existing work distinguishes between two regimes, depending on the free energy barrier presented to monomers binding on the surface: chemisorption if the reaction barrier is high, and physisorption if the barrier is low and the characteristic time to establish a bond on contact is short.
In chemisorption, the reaction time of the monomers is larger than the time for the polymer to return to an equilibrated state, so the process becomes quasistatic. Theory~\cite{OShaughnessy2003,Panja2009}, experiment~\cite{OShaughnessy2005}, and simulations~\cite{Shaffer1993,Shaffer1994,Ponomarev2000,OShaughnessy2005,Panja2009}, have all shown that two mechanisms control chemisorption: the zipping down of sequential monomers, and the formation of extra nucleation points via loop formation. Loop formation lowers the adsorption time relative to a simple sequential zipping mechanism, and so chemisorption is said to take place via the accelerated zipping mechanism.

Physisorption is, on average, a simple zipping mechanism: sequential monomers very quickly attach to the surface, leaving no time for loop formation~\cite{Ponomarev2000,Descas2006,Bhattacharya2008,Panja2009}. This forces the chain out of equilibrium, as the remaining unabsorbed segment initially moves more slowly than the chain zipping down. The precise scaling of this adsorption time depends on the strength of the polymer-surface attraction~\cite{Panja2009}. For irreversible physisorption, the intermediate chain conformation combining  a stretched tether at the zipping end, and a coil at the free end, is known as the `stem and flower' model, and was first described by Brochard-Wyart~\cite{Brochard1995} for tethered polymers under strong shear flow.

Many of the systems considered are entirely homogeneous polymer chains with no specificity of binding sites, while others examined copolymers attaching to uniform surfaces~\cite{Shaffer1994,Milchev2011}. Li et al. have studied stripe-patterned surfaces~\cite{Li2015}, and copolymers of one attractive and one inert monomer type~\cite{Li2016}, but many processes in self-assembly are much more specific than this.  These processes, such as DNA hairpin formation \cite{Libhaber1998,Dudko2007}, still show zipper kinetics.

In this paper, we examine a trade-off in the entropy barrier faced by reaching the distant target against the reduction in confinement. If we make the chain very short, there will be a very small chance of it stretching far enough to reach the receptor site, and the time to reach the receptor will be long. If we make the chain very long, there will no longer be such an entropic penalty for reaching the same receptor. However, the chain will now be able to explore a very large volume, and that reduces the probability that the binding site will hit the target. Once the expression for the average binding time $\tau_\mathrm{on}$ is obtained, we find the optimal chain length for the fastest binding time: this turns out to be exactly when the target separation $a$ is equal to the radius of gyration of the chain. Finally, we consider the time of chain adsorption to a sequence of receptors, calculating it as a function of chain length, number of binding sites and the distance between them. 

This process of adsorption may proceed via different pathways, involving purely sequential (zipper) single steps, or multiple-distance looping events. It turns out that,  for sufficiently separated binding sites, the simple zipping mechanism becomes the preferred pathway. We  also consider the `stem and flower' effect, and show that for a chain looking to bind its free end to a target a distance $a$ away as fast as possible, non-equilibrium effects associated with a slow drift of the chain towards the receptor targets defines a certain optimal number of intermediate receptors that achieves the fastest mean adsorption time.

\section*{Diffusion of a grafted ligand}

We consider an ideal polymer chain ($N$ segments of length $b$) that is grafted at the origin to a flat surface, where the last ($N$th) monomer is the binding ligand. 
To find the equilibrium distribution of the chain configuration, we use the Gaussian chain propagator of an ideal chain $G_N(\mathbf{r},\mathbf{r}_0)$. This is the probability that such a chain begins at $\mathbf{r}_0$ and ends at $\mathbf{r}$, and it satisfies the Edwards equation~\cite{DoiEdwards}:
\begin{equation}
\left(\frac{\partial}{\partial N}-\frac{b^2}{6}\nabla^2_{\mathbf{r}}+\frac{U_{e}(\mathbf{r})}{k_{B}T}\right)G_N(\mathbf{r},\mathbf{r}_0)=\delta(\mathbf{r}-\mathbf{r}_0)\delta(N)  \label{eq:edw}
\end{equation}
The potential $U_e(\mathbf{r})$ represents any external forces applied to the monomer, and we here will consider the basic problem with $U_e = 0$ . We need to implement a boundary condition on the substrate plane $z=0$. This is a question with a very long history \cite{DiMarzio}, culminating with the classical work of Edwards and Freed \cite{Edwards1969} about the `chain in a box'. Many aspects of this problem, of a chain near a hard wall, were explored over the years, with seminal contributions \cite{Joanny1979,DeGennes1980,Bickel2001} being just a few of many important references, all using and exploiting the `exclusion' condition:
\begin{equation}\label{eq:edbc}
G_N(\mathbf{r},\mathbf{r}_0) \big|_\mathrm{surface}=0. 
\end{equation}
This means, in the case of \eqref{eq:edw}, that no monomer may rest against the wall. Surprisingly, this restriction is not well covered in the literature, and it is difficult to acquire intuition for it. Exclusion seems drastically different from the (correct) reflecting boundary condition we impose on Brownian particles, if they were not connected on the chain. This is a subtle, yet potent effect of chain configurational entropy -- understood first by DiMarzio from the point of view of counting restricted  chain configurations \cite{DiMarzio}, and then by Edwards and Freed by looking at the entropic repulsive force arising if we were to push the chain into a wall \cite{Edwards1969}.  

When only one planar wall is present, the solution of the Edwards equation (\ref{eq:edw}) involves one negative chain image. 
Although we tether the chain at the origin $\mathbf{r}=0$, it is necessary to insist that the first monomer steps directly away from the surface, so $z_0=b$, and the image chain starts at $\bar{z}_0=-b$. The remaining chain is then of length $N-1$, but since we must assume $N$ is large, we ignore this:
\begin{equation}\label{eq:propend}
G_{N}(\mathbf{r})=\left(\frac{3}{2\pi Nb^2}\right)^{3/2}e^{-\frac{3(x^2+y^2)}{2Nb^2}}\left[e^{-\frac{3(z-b)^2}{2Nb^2}}-e^{-\frac{3(z+b)^2}{2Nb^2}}\right].
\end{equation}

\begin{figure}[b]
	\centering
	\includegraphics[width=0.48\textwidth]{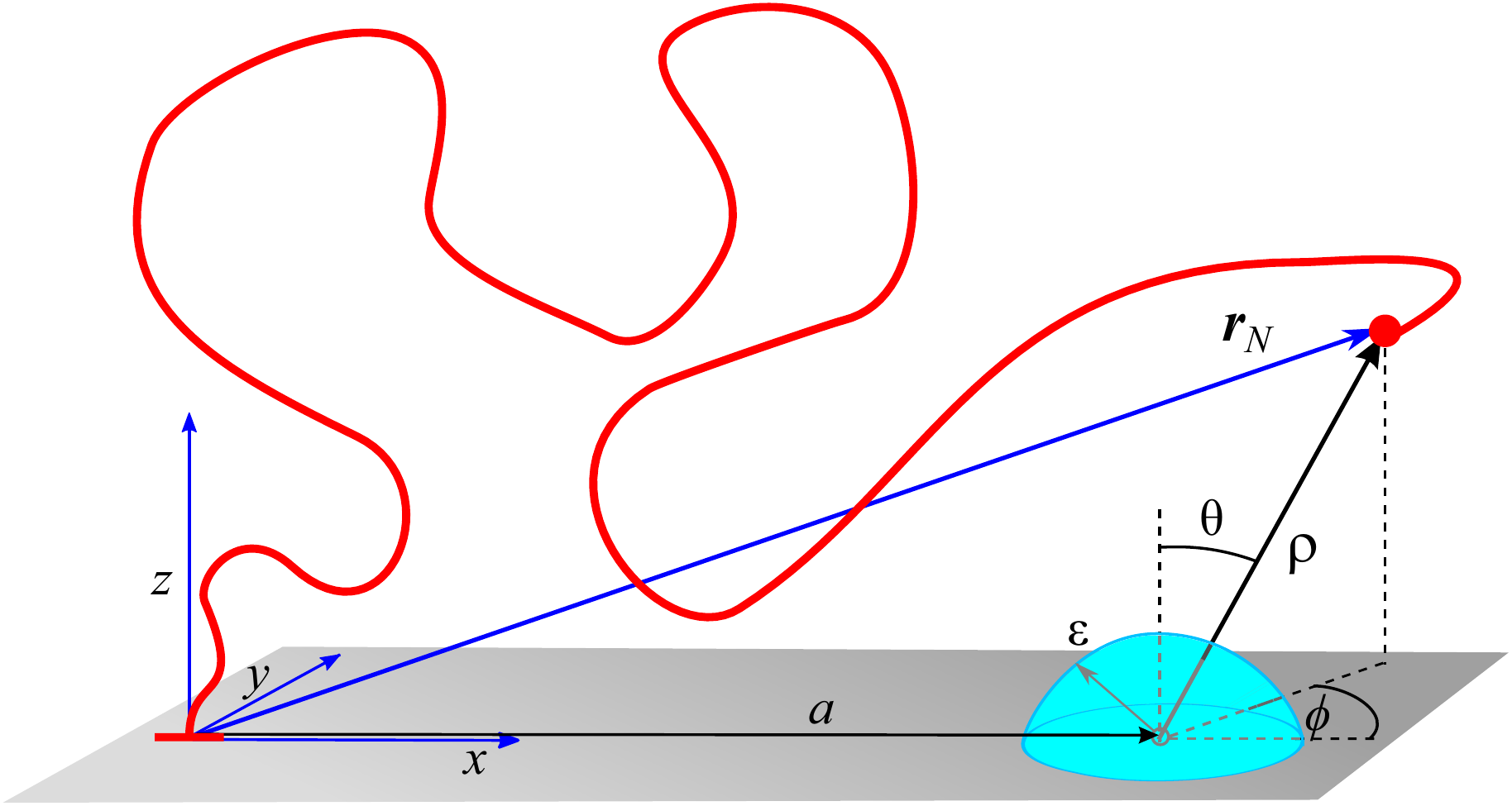}
	\caption{A Gaussian chain of $N$ monomers is tethered to a hard wall at the origin. A hemispherical absorbing target of radius $\varepsilon$ lies on the surface, a distance $a$ from the tether. We switch from Cartesian coordinates about the tether to spherical polar coordinates about the target, to find the mean first passage time of the end of the chain onto the target.}
	\label{fig:schem}
\end{figure}

 The binding ligand (the $N$th monomer) needs to find a surface receptor placed a distance $a$ from the grafting site, as illustrated in Fig.~\ref{fig:schem}. The receptor zone is assumed hemispherical, with a small radius $\varepsilon$. We will now construct an effective  radial probability distribution for the  distance $\rho$ from the binding site $\bm{r}_N$ to the target receptor. That radial probability distribution $P_{eq}(\rho)$ becomes amenable to the first passage time approach of Szabo et al.~\cite{Szabo1980}.

In Eq. (\ref{eq:propend}), the propagator for the position of the chain end is presented using a Cartesian coordinate system with the origin at the point of grafting. However, since we are looking for the passage time into a hemisphere centered on a receptor, it useful to shift the to spherical polar coordinates $(\rho, \theta, \phi)$ centered on the target,  Fig.~\ref{fig:schem}. Then we will need to integrate over the two angles,  to finally derive the radial probability density about the target receptor, $P(\rho)$, which will be a function of the receptor position $a$. 
Let us choose the target to be in the positive $x$-direction relative to the tethered end. Thus we can write the coordinate transformation as
\begin{equation}
x - a = \rho\sin\theta\cos\phi\,, \;  
y = \rho\sin\theta\sin\phi\, , \; z = \rho\cos\theta.
\end{equation}
The two scalar products in the combined exponents of Eq.(\ref{eq:propend}) become:
\begin{equation}
(\mathbf{r}\pm b\hat{\mathbf{z}})^2=a^2+b^2+\rho^2+2a\rho\sin\theta\cos\phi\pm2b\rho\cos\theta .
\end{equation}

The next step of integration over the solid angle on the unit hemisphere is not easy. We need to evaluate
\begin{equation} \label{eq:I}
I = \int_{0}^{\pi/2}d\theta\sin\theta\int_{0}^{2\pi}d\phi \;e^{-\alpha\cos\phi\sin\theta\pm\beta\cos\theta}, 
\end{equation}
where parameters $\alpha$ and $\beta$ involve $N,b,a$ and $\rho$.
This is solved by realizing that the integrand has a non-trivial axial symmetry. To exploit this symmetry, one needs to transform back into Cartesian coordinates about the target: $x'=\cos\phi \sin\theta$, $z'=\cos\theta$, and rotate these new coordinates by an angle $\varphi=\tan^{-1}(\beta/\alpha)=\tan^{-1}(b/a)$ around the $y$-axis. The direction of this rotation depends on the sign of the $z$-term in the exponent (i.e. whether we are dealing with the `real' or `image' Gaussian term). The details of this calculation are given in Supplementary part \ref{app:A}, including the full expression for the normalized radial distribution function $P_{eq}(\rho)$. It turns out that a very good approximation exists to that complicated expression, which is nearly accurate except for the region $a\leq b$. We believe it is safe to assume the receptor site can never be this close to the grafting site, and proceed with the useful approximation:
\begin{align}\label{eq:prad}
&P_{eq}(\rho) \approx \frac{2b}{a} \sqrt{\frac{N\pi}{6}} \left(\frac{3}{2\pi Nb^2}\right)^{3/2} e^{-\frac{3(a^2+\rho^2)}{2Nb^2}} I_{1}\left(\frac{3a \rho}{Nb^2}\right), 
\end{align}
where $ I_{1}(..)$ is the  the 1st rank modified Bessel function of the first kind, and $\sqrt{6/N\pi}$ is the thermodynamic partition function of a grafted Gaussian chain.

\begin{figure}
	\centering
	\includegraphics[width=0.45\textwidth]{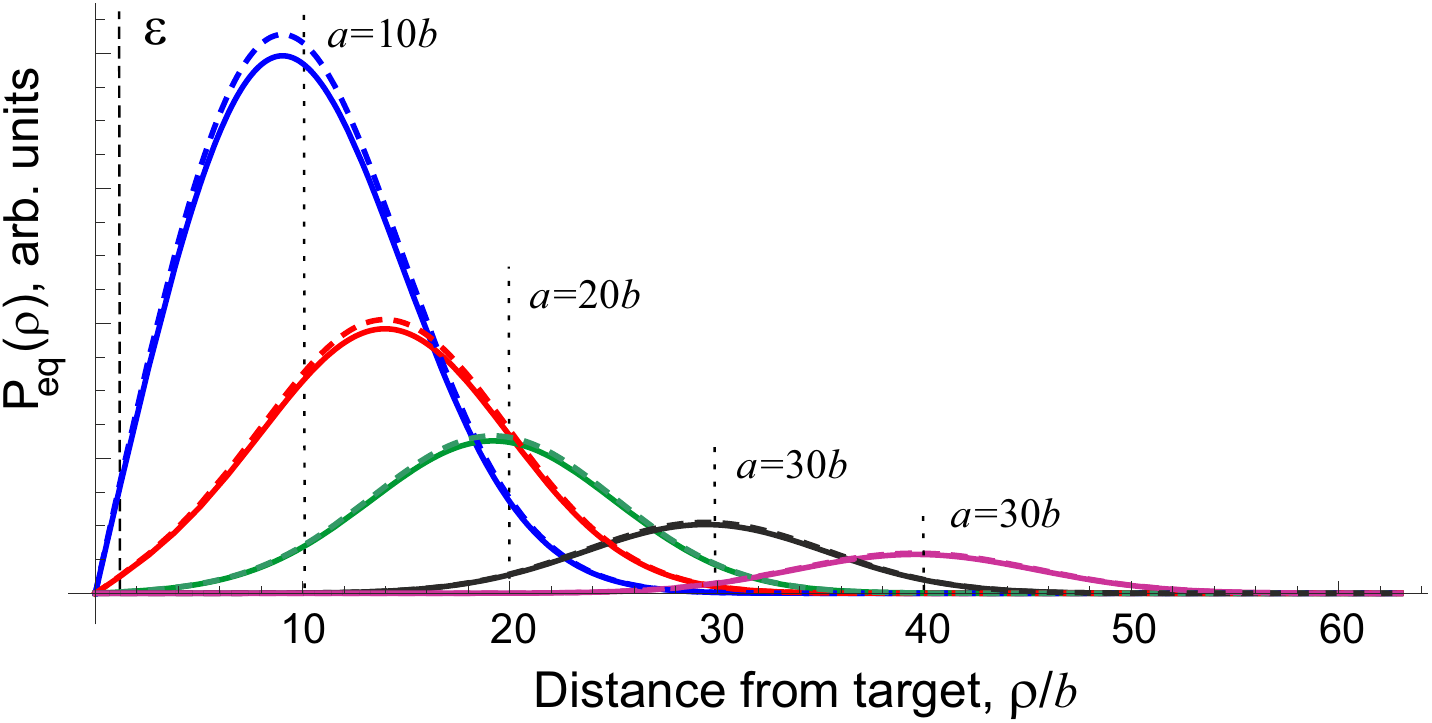}
	\caption{The radial probability density $P_{eq}(\rho)$ of the end of the chain about the hemispherical target, given by the approximation \eqref{eq:prad}, for $N=100$ (so $R_g=10b$ here), and several values the target positions: $a/b=10, 15, 20, 30$ and $40$. Dashed lines show the exact result of the angular integration (\ref{eq:I}) for comparison; the deviations are only seen at small $a$. }
	\label{fig:prad}
\end{figure}

This distribution is plotted in Fig.~\ref{fig:prad}. The probability density goes to zero as $\rho\to 0$ because of exclusion at the surface, and then peaks before decaying away again. This distribution peak is going to be close to the target distance: $\rho \approx a$, because the chain is most likely to be found near the tether. The actual peak lies at just less than $\rho=a$, as a result of averaging over the polar angles, but the difference becomes less significant as $a \gg R_g$.

\begin{figure}
	\centering
	\includegraphics[width=0.33\textwidth]{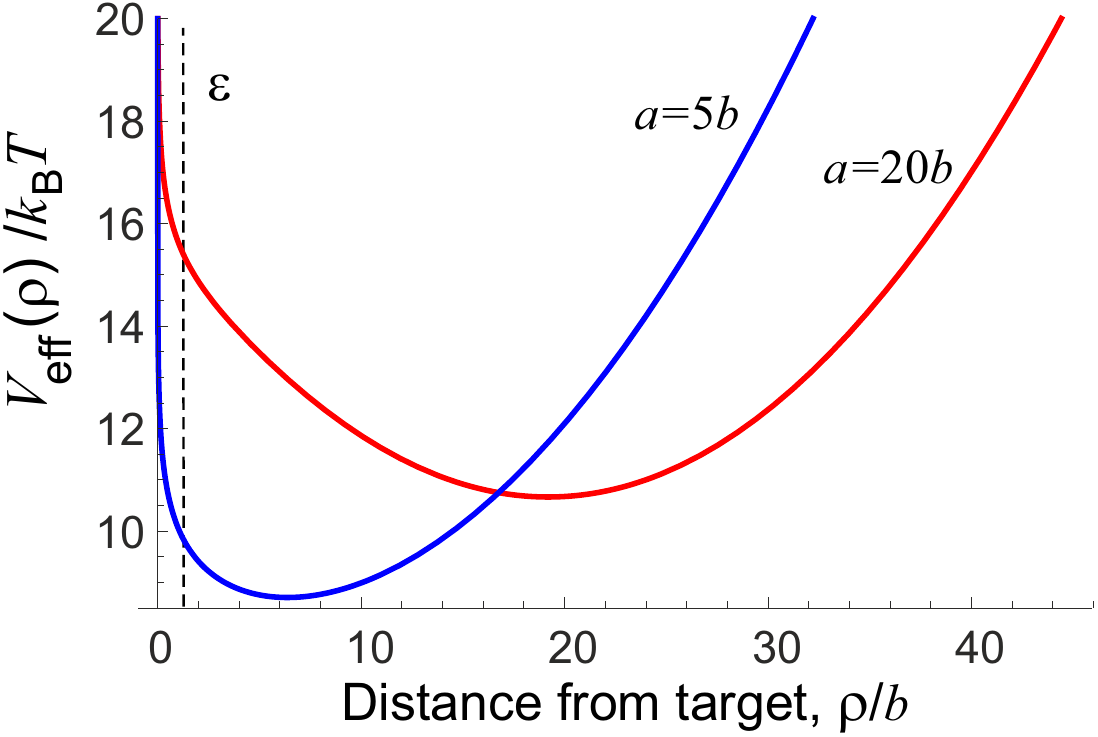}
	\caption{The effective potential $V_\mathrm{eff}(\rho)$ of the end of the chain, plotted for $N=100$ (so $R_g=10b$ here), and two target positions: $a/b=5$ and $20$. The example of the target size $\varepsilon = 1.5b$ is marked with a dashed line. }
	\label{fig:Veff}
\end{figure}

One can identify the radial probability density discussed above with an effective radial potential via the Boltzmann factor: $V_{\mathrm{eff}} = -k_BT \, \ln \, P_{eq}(\rho)$. The resulting effective potential that the binding ligand on the $N$th chain segment experiences is a function of distance from the target receptor, and depends on two relevant length scales in the problem: the chain radius of gyration $R_g=N^{1/2}b$, and the distance to target $a$. It is plotted in Fig. \ref{fig:Veff} for two values of $a$: one above and one below the $R_g$. The effective potential has a minimum (seen as the peak of the radial probability distribution), but diverges in the close proximity to the target because of the exclusion boundary condition the wall imposes on the chain: this produces an effective (entropic) repulsion that the ligand has to overcome to reach the target at $\rho \to 0$. We see in Fig. \ref{fig:Veff} that this effective energy barrier (between the minimum of $V_\mathrm{eff}$ and the value at $\rho=\varepsilon$) is $\sim 2k_BT$ for $a=5b$, raising to $\sim 5k_BT$ for $a=20b$, for the chain of 100 monomers.

\section*{Mean first passage time to target}
Once we have the equilibrium probability distribution for the single radial variable $\rho$ (the distance of the dangling ligand from the target receptor), we can use a famous relation derived by Szabo et al.~\cite{Szabo1980} to find the mean first passage time  (MFPT) to an absorbing surface at $\rho =\varepsilon$. This time is obtained by evaluation:
\begin{equation} \label{eq:szabomfpt}
\tau = 2\pi \int_{\varepsilon}^{\infty}d\rho\left[D \rho^2 P_{eq}(\rho)\right]^{-1}\left[\int_{\rho}^{\infty}d\rho'\; \rho'^2 P_{eq}(\rho')\right]^2 , 
\end{equation}
where we assume the diffusion coefficient of the free end of the chain, $D=k_{B}T/\gamma$, is constant and equal to the diffusion coefficient of a single monomer in solution. In the free particle case, it is necessary to constrain the particle with an upper reflective boundary, but in our case of a Gaussian polymer chain, the entropic spring effect ensures the integrals converge if we take the upper limit to $\infty$.

Even for the approximate probability distribution given in \eqref{eq:prad}, the integral in \eqref{eq:szabomfpt} does not have an easy analytical solution. However, much progress can be achieved by noticing that the integrand diverges as $\rho\to 0$, and so the main contribution to the mean first passage time comes from the region of small $\rho$. The technical details of this calculation can be found in the Supplementary part \ref{app:B}. Expanding the integrand about $\rho=0$ and retaining only the leading term, we find that \eqref{eq:szabomfpt} reduces to a simple integral
\begin{align}\label{eq:mfpt}
\tau_\mathrm{on} &\approx \frac{2 N^2b^4}{9 D}e^{3a^2/2Nb^2}\int_{\varepsilon}^{\infty}\frac{d\rho}{\rho^3} =
 \frac{N^2b^4}{9D\varepsilon^2}e^{3a^2/2Nb^2},
\end{align}
where, as before, we recognize the characteristic length scale $R_g=N^{1/2}b$, the radius of gyration of an ideal chain.

\begin{figure}
	\centering
	\includegraphics[width=0.4\textwidth]{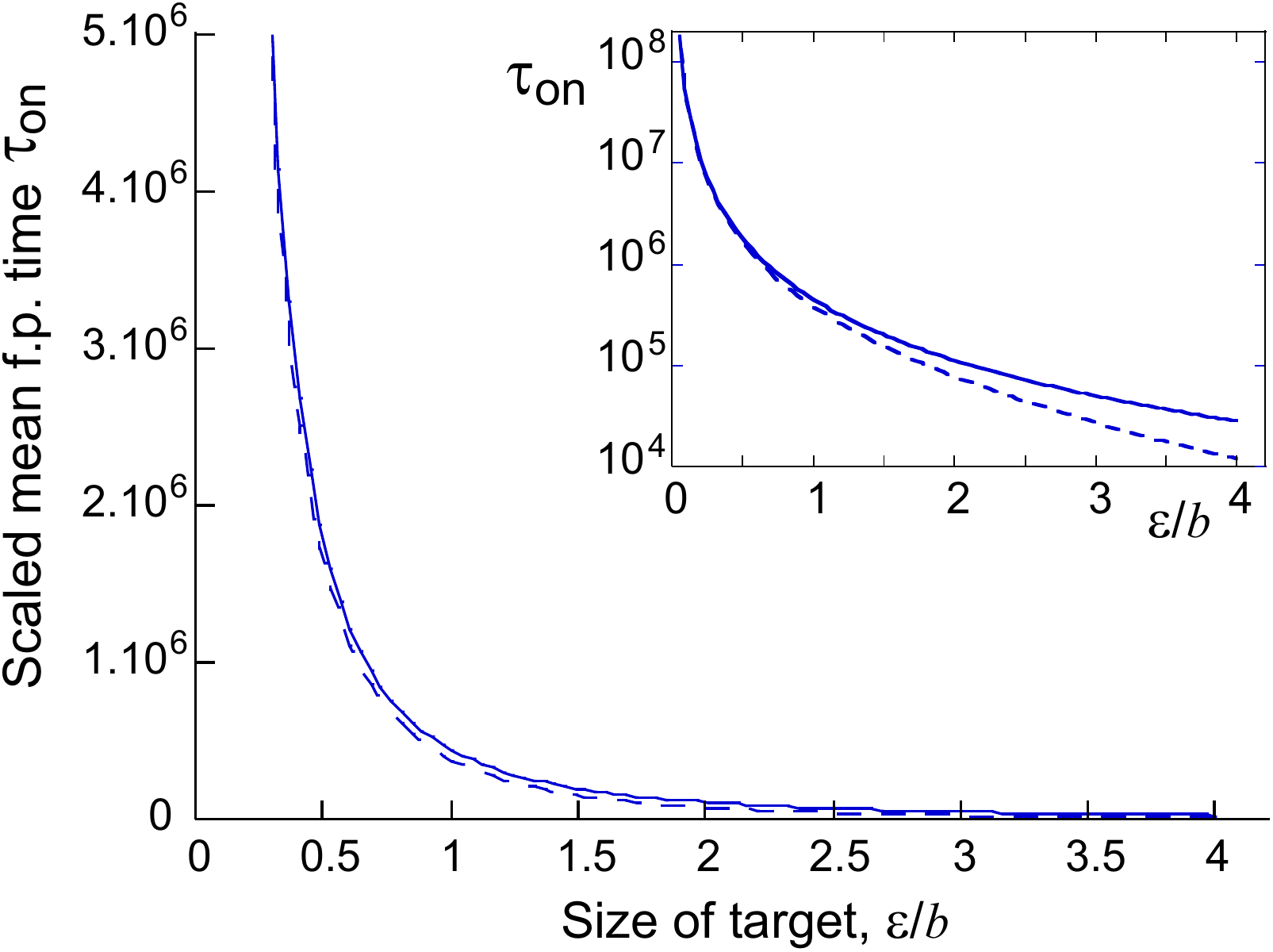}
	\caption{The mean first passage time $\tau_\mathrm{on}$ (in units of $b^2/D$) calculated numerically from \eqref{eq:szabomfpt} (dashed line), compared with the approximation of \eqref{eq:mfpt} (solid line). Here, again, $N=100$, and $a=20b$. In reality, one might imagine the relevant size of targets for specific binding to be not much greater than $b$.  The inset shows the same plot on logarithmic scale, which covers a greater range and also emphasizes the deviations of the approximate expression for $\tau_\mathrm{on}$ at the tail of the function.}	\label{fig:mfpt}
\end{figure}

Equation (\ref{eq:mfpt}) is the first of two main results of this paper.
The comparison of the exact numerical integral and the approximation in \eqref{eq:mfpt} above is shown in Fig.~\ref{fig:mfpt}, where the mean time of the binding ligand reaching the target receptor is plotted against the `size' of the receptor (measured by the radius of the hemisphere $\varepsilon$, see the sketch in Fig. \ref{fig:schem}). The deviations are enhanced in Fig. \ref{fig:mfpt} inset by the logarithmic scale, and are evidently very small for sufficiently small targets. 

Clearly, \eqref{eq:mfpt} is a good approximation, offering a compact analytical expression that we can examine. One can compare it with the average time for a free  polymer chain to make a loop by having the last $N$th monomer reach a sphere of radius $\varepsilon$ around the first monomer \cite{Szabo1980} (the Szabo problem, corresponding to distance $a=0$ in our case, and no restricting surface -- solved using our method in Supplementary part \ref{app:B}):

\begin{equation} \label{eq:sza}
\tau_\mathrm{loop}=\sqrt{\frac{\pi}{54}} \frac{(Nb^2)^{3/2}}{D\varepsilon} . 
\end{equation}
Also instructive is to compare our expression (\ref{eq:mfpt}) with the average time for a free Brownian particle to escape a closed volume $V$ through a small hole of size $\varepsilon$ \cite{Schuss2007} (the `narrow escape problem' of Holcman et al.), which is estimated as  $\tau_\mathrm{esc} = V/D\varepsilon$. If the volume is replaced by the average extent of chain spreading, $V=R_g^3$, this matches the Szabo expression (\ref{eq:sza}). Both have a different scaling with the size of target: $1/\varepsilon$ compared to $1/\varepsilon^{2}$ in \eqref{eq:mfpt}. In our case the chain is strongly inhibited from approaching the wall due to the polymer-specific boundary condition; as a result the average time it takes to reach the target is much longer even without the additional exponential factor reflecting the entropic barrier for binding. 

The second factor that distinguishes the mean binding time  in \eqref{eq:mfpt} is the exponential factor $\exp[3a^2/2R_g^2]$. This represents thermal activation over an entropic barrier $\Delta G = \frac{3}{2} k_BT a^2/Nb^2$, which is essentually the free energy to stretch the chain ends by a distance $a$.  This factor, significantly increasing the time for bridging to a distant target, only arises for the tethered chain (all polymer work on the related narrow escape problems~\cite{Muthukumar1988,Muthukumar1999,Szabo2000}  has thus far focused on polymers with no attachment to the boundary of the domain, which fundamentally alters the accessibility of the binding site).

\section*{Bridging across a gap}

During cell-cell adhesion, the two cell membranes come very close to one another. Thus, they can be modelled as two parallel planes a distance $d$ apart, with their actual curvature playing a minor role in the dynamics. As in the single plane case of the previous sections, we must consider the first monomer as stepping directly away from the surface, so the tether in this coordinate system is at (0,0,b). We can then write down the chain propagator in exactly the manner of Edwards and Freed \cite{Edwards1969}, separating the unconstrained chain in the $xy$-plane from the narrow confining box along $z$, with one chain end fixed at $z=b$:
\begin{align} \label{eq:propEdw}
G_N(x,y,z)=\frac{2}{d}\sum_{n=1}^{\infty}&\sin\left(\frac{n\pi z}{d}\right)\sin\left(\frac{n\pi b}{d}\right)e^{-\frac{n^2\pi^2 Nb^2}{6d^2}} \nonumber \\
&\times\frac{3}{2\pi Nb^2}e^{-\frac{3(x^2+y^2)}{2Nb^2}}.
\end{align}
As before, both planes are monomer excluding due to the chain entropic repulsion. When cells are close to one another, the distance $d^2\ll Nb^2$, we are free to consider only the first term in the sum, as the exponential in the sum suppresses subsequent terms (a  regime known as the ground-state dominance in polymer physics). 

As in the single-plane case, the key is to derive the radial probability distribution $P_{eq}(\rho)$ for the ligand a distance $\rho$ away from the binding site, see Fig. \ref{fig:2cells}. If the target receptor is placed a perpendicular distance $a$ from the tether, on the opposite plane, then there is no obvious symmetry to exploit. Instead, it is possible to make progress if we use the approximation $a\gg\varepsilon$. The propagator is radial in the $xy$-plane about the tether. If the receptor is placed in the $x$-direction, then around the receptor, at small $\rho$, the gradient of the propagator will have no $y$-component to first order. Therefore we can assume that $y=0$ in the propagator without significantly changing its value. This allows us to build our hemispherical shells centered on the target  by shifting the coordinate system by $x'=x-a$, then integrating the propagator over semicircles of radius $\sqrt{\rho^2-x'^2}$, holding $x'$ constant. This eliminates the $z$-dependence, effectively generating the average $\langle\sin \left(\pi z/d\right)\rangle$. The second integration is over the $x'$-axis from $-\rho$ to $+\rho$, adding these semicircles with an appropriate surface element to recover the radial distribution function about the receptor. This distribution $P_{eq}(\rho)$ can be expanded at small $\rho$ again, exploiting the vanishing denominator as in Eq. \eqref{eq:mfpt}, and in the same way we obtain the result for the mean first passage time:

\begin{figure}[t]
	\centering
	\includegraphics[width=0.47\textwidth]{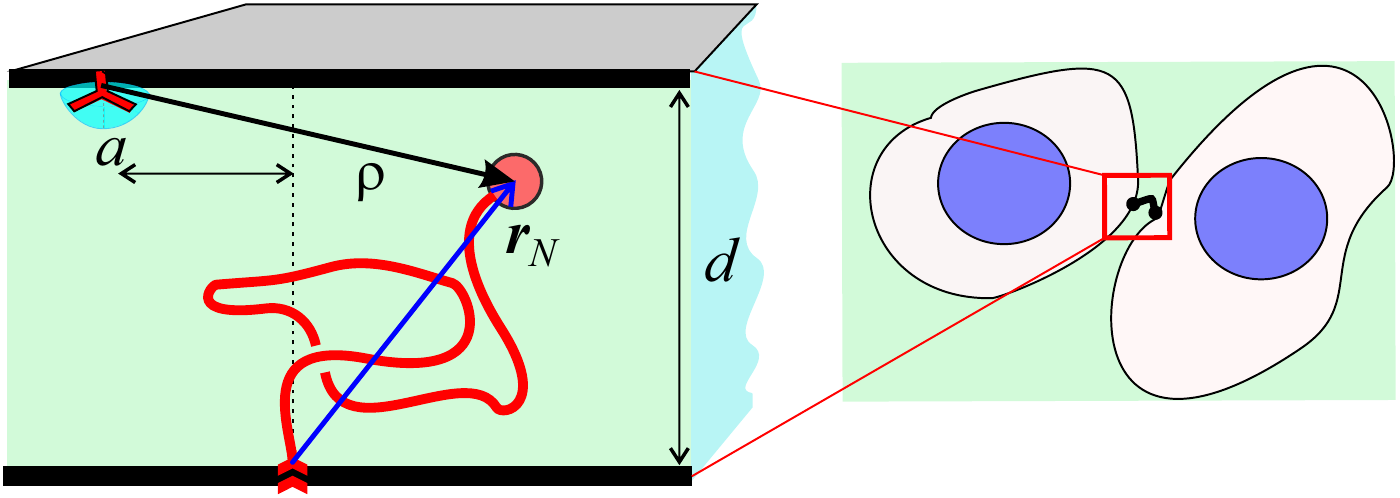}
	\caption{A schematic for the cell-adhesion process, where a flexible linker will bind to a neighbouring cell across a gap $d$. The inset shows our simplified scenario: assuming the two membranes to be infinite in extent, and parallel to each other, with the receptor displaced by a distance $a$.}\label{fig:2cells}
\end{figure}

\begin{equation}\label{eq:paraclose}
\tau=\frac{2Nb^2d^2}{3\pi^2D\varepsilon^2}e^{3a^2/2Nb^2}.
\end{equation}
  
One can see the same constrained dependence on the reaction volume $\varepsilon$, due to the difficulty for any chain segment to get that close to the wall. It is possible to consider the $a=0$ case using a different method, relying on the azimuthal symmetry of the propagator. We recover the non-exponential prefactor in Eq. \eqref{eq:paraclose}, confirming the validity of the analysis. It is also possible to adapt this result for a very small gap, $d<\varepsilon$, where the 2D scaling relation $\tau\sim (Nb^2/D)\ln(\sqrt{Nb^2}/\varepsilon)$ replaces Eq.~\eqref{eq:paraclose}. See Schuss et al.~\cite{Schuss2007} for a freely diffusing Brownian particle analogue.

The quadratic dependence of the mean bridging time on the cell gap $d$ in Eq.~\eqref{eq:paraclose} is the novel feature, but it is only valid in the tightly confined case $d \ll R_g$. The general expression for the binding time is complicated, but the calculation in the opposite limit (short chain or wide gap: $d^2\gg Nb^2$) is presented in the Supplementary part C. We find a very accurate interpolation formula for the mean time of bridging between two surfaces, which spans across the whole range between the two limiting cases:
\begin{equation}\label{eq:tau2}
\tau_2 =\frac{2R_g^2d^2}{3\pi^2D\varepsilon^2}\frac{1}{(1+36 d^4/\pi^2R_g^4)}\cdot e^{3(d^2+a^2)/2R_g^2} \ .
\end{equation}

\section*{Accelerated zipper binding}

We can now use the expression \eqref{eq:mfpt} for $\tau_\mathrm{on}(a,N)$ to examine a simplified version of the multiple-site binding problem. Let us consider $M$ binding sites spaced evenly along a polymer chain of total length $N$ (so that the ligands are $\Delta N = N/M$ monomers apart, along the chain). As in the single-site problem, we take the first segment of the chain to be already bound (grafted) to the surface, so there are $M$ binding events yet to occur in total. The receptors for these ligands are spaced at equal distances $\Delta a$ apart in a straight line on a plane reflecting surface, see Fig. \ref{fig:binding}(b). While it should not be difficult to consider arbitrary positioning of chain binding sites and surface receptors, we use this simplified geometry in the hope of finding a clear analytical result for the average binding time.

Assuming that each binding ligand on the chain associates with a specific receptor, the chain may form a loop by binding across several receptors,  a distance $q \Delta a$ from the grafting point, see Fig. \ref{fig:binding}(b) for a $q=2$ loop.  The time to bind to a receptor a distance $q \Delta a$ away is
\begin{equation} \label{eq:taum}
\tau_{qa}=\frac{(qN/M)^2b^4}{9D\varepsilon^2}e^{\frac{3(q \Delta a)^2M}{2qNb^2}}  =  q^2 e^{\frac{3(q-1)M \Delta a^2}{2Nb^2}} \cdot \tau_{1a},
\end{equation}
where the single-step binding time $\tau_{1a} = \tau_\mathrm{on}(\Delta a, N/M)$ from Eq. \eqref{eq:mfpt}.  If the chain does not bind sequentially, but with a loop forming (for example, next nearest binding in Fig.~\ref{fig:binding}(b)), the subsequent binding of the `middle'  ligands is much faster (site [C] between [B] and [D] in Fig.~\ref{fig:binding}(b)), and therefore is not a rate-limiting step. The combination of single and multiple steps is the accelerated zipper mechanism.

The kinetics of both single and multiple steps can be understood as a continuous-time Markov chain~\cite{Medhi1994,Norris1998}, with $M+1$ discrete states corresponding to how far along the chain the final binding event has been. This means we can write the rate equations in vector form: $\mathrm{d}P/\mathrm{d}t = Q \cdot P$, where $Q$ is known as the rate matrix. This is called the backward Kolmogorov equation. The $(M+1)\times (M+1)$ rate matrix $Q$ has the following form:
\begin{equation}
\begin{pmatrix}
-\sum_{q=1}^{M}k_q & k_1 & k_2 & k_3 & ... & k_M \\
0 & -\sum_{q=1}^{M-1}k_q & k_1 & k_2 & ... & k_{M-1} \\
0 & 0 & \ddots & \ddots & \ddots & \vdots \\
\vdots & ... & 0 & -\sum_{q=1}^{2}k_q &k_1 &k_2 \\
\vdots & 0 & ... &0 & -k_1 & k_1 \\
0 & 0 & 0 & ... & 0 & 0 \\
\end{pmatrix}
\end{equation}
where $k_q=1/\tau_{qa}$ are the rates of binding to a receptor a distance $q\Delta a$ away. Instead of explicitly solving the  Kolmogorov equation, we can rely on the following fundamental result to derive the recursive relations for mean first passage times $\langle\tau(M+1-i)\rangle$ from state $i$ to the final fully bound state (across $M+1-i$ receptors, where state $i=0$ is the tethered chain with no receptors bound, and $i=M+1$ is when the final $M$th receptor is bound)\cite{Norris1998}:
\begin{equation}\label{eq:Qexp}
\sum_{j} Q_{ij}\langle\tau(M+1-j)\rangle = -1 \ ,
\end{equation}
for all states $1\leq i< M+1$. If we start in the final absorbing state, then the mean first passage time is zero by definition, and so $\langle\tau(0)\rangle=0$. The remaining $\langle\tau(i)\rangle$ can then be constructed recursively.

\begin{figure}[t]
	\centering
	\includegraphics[width=0.4\textwidth]{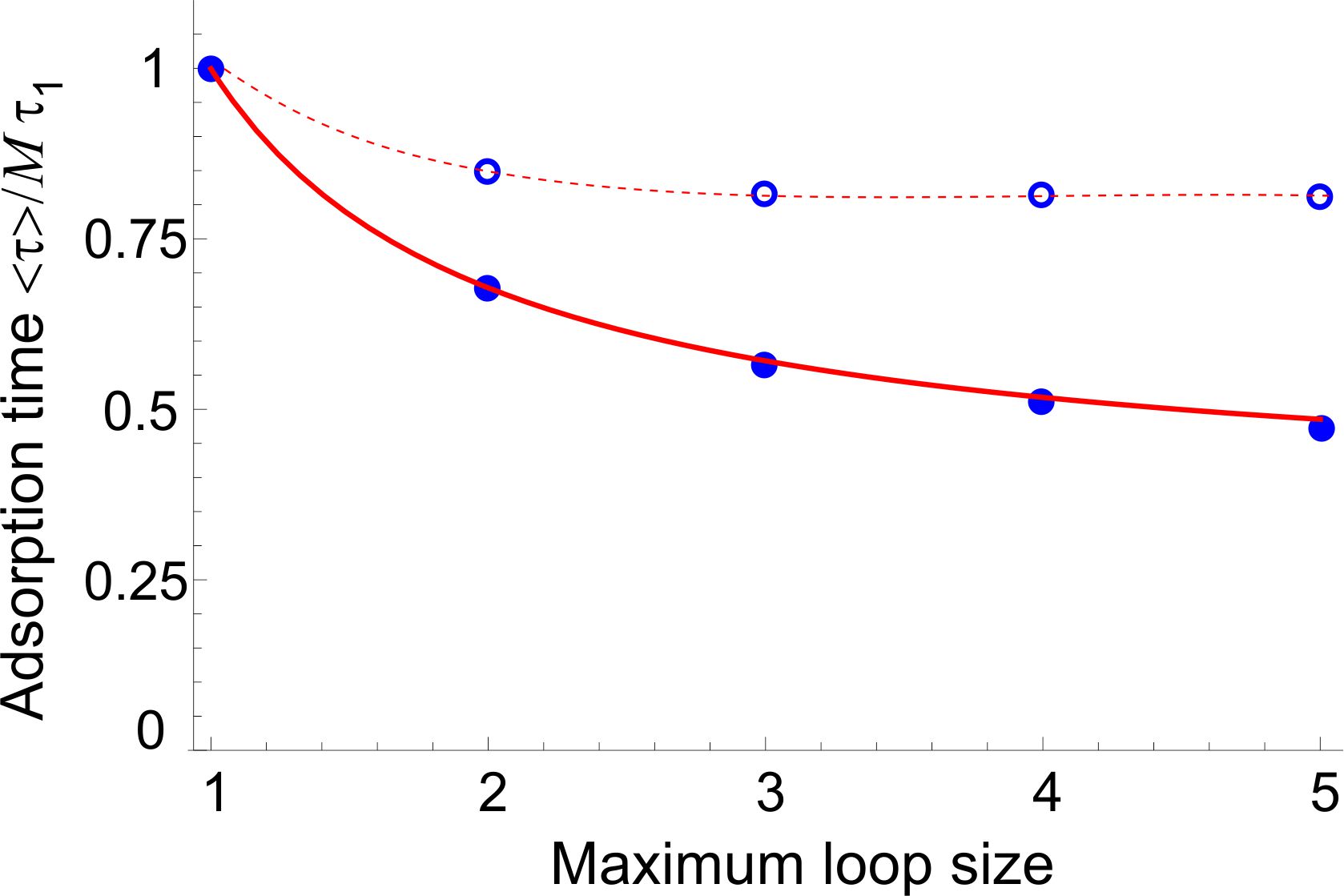}
	\caption{The plot of the (scaled) average adsorption time $\langle \tau(M) \rangle$ for $M=20$, when the multiple-step jumps (loops) are allowed. The $x$-axis indicates the largest jump allowed to accelerate the zipper binding. Filled symbols $\bullet$ represent the maximum effect, when the exponential entropic-penalty factor in $\tau_{1a}$ is equal to one (i.e. the distance $\Delta a=0$); the open symbols $\circ$ represent the reduced acceleration effect when $\Delta a^2=\Delta N \, b^2$.  }
	\label{fig:loops}
\end{figure}

While Eq. \eqref{eq:Qexp} is a full solution to the problem of mean binding time, it does not indicate how important loop formation is in the binding process. To make progress, we compare the binding times $\tau_{qa}$. For multiple steps, the rapidly increasing entropic barrier to reach receptors further away means that the expected time for binding gets longer and longer. Figure~\ref{fig:loops} shows how the mean binding time across $M=20$ receptors reduces as we include the possibility of longer jumps. It is clear that we are free to neglect steps past $q=2$ when $\Delta a^2 \geq \Delta N b^2$ (the open symbols in Fig.~\ref{fig:loops}, as $\tau_{qa}$ rapidly increases with $q$, and the adsorption time seems to rapidly approach a limiting value. For very closely spaced receptors ($\Delta a^2 \ll (N/M)b^2$) the ratio of  $\tau_{qa}/\tau_{1a} =  q^2$, and Fig.~\ref{fig:loops} shows that although the largest reduction in binding time comes from increasing the maximum step to $q=2$, larger size loops do still play an appreciable role. To offer a quantitative idea of how big a role they play, we manually fitted a curve to the closely spaced receptors in Fig.~\ref{fig:loops} (the solid red line), and found that the deviation from a fitted limiting value $\langle\tau_\infty\rangle$ was $\langle \tau_q\rangle-\langle\tau_\infty\rangle\propto 1/q$.

In the regime $\Delta a^2\geq \Delta Nb^2$, we restrict the binding process to either binding at the closest available site, which takes an average time $\tau_{1a}$ and follows in a zipper sequence, or at the next nearest site, which takes $\tau_{2a}$, as shown in Fig.~\ref{fig:binding}(b); all other binding events across greater distances are neglected ($k_i=0$ for $i>2$). Then, Eq. \eqref{eq:Qexp} defines a recurrence relation for arbitrary $M$:
\begin{align}
&\langle\tau(1)\rangle = \frac{1}{k_1}, \quad \langle\tau(2)\rangle = \frac{2}{k_1+k_2}, \\ 
& (k_1+k_2)\langle\tau(M)\rangle = 1 + k_1\langle\tau(M-1)\rangle+k_2\langle\tau(M-2)\rangle   \nonumber \\
& \mathrm{for}\quad M\geq 2. \nonumber
\end{align}
Using a standard generating function method to solve the recurrence relation~\cite{Wilf1990}, we find that
\begin{equation} \label{eq:taufull}
\langle\tau(M)\rangle = \frac{Mk_1^2 +2(M+1)k_1k_2+2k_2^2\left(1+\left(-\frac{k_2}{k_1+k_2}\right)^{M-1}\right)}{k_1(k_1+2k_2)^2}.
\end{equation}
Since this expression is only valid for sufficient spacing, $k_2\ll k_1$, and we are free to Taylor expand Eq. \eqref{eq:taufull}:
\begin{align}\label{eq:tauapp}
\langle\tau(1)\rangle &\approx M\tau_{1a}-\frac{2(M-1)\tau^2_{1a}}{\tau_{2a}}\nonumber \\
&=\left(M-\frac{M-1}{2}e^{-\frac{3M\Delta a^2}{2Nb^2}}\right)\tau_{1a}.
\end{align}

\begin{figure}
    \centering
   \includegraphics[width=0.45\textwidth]{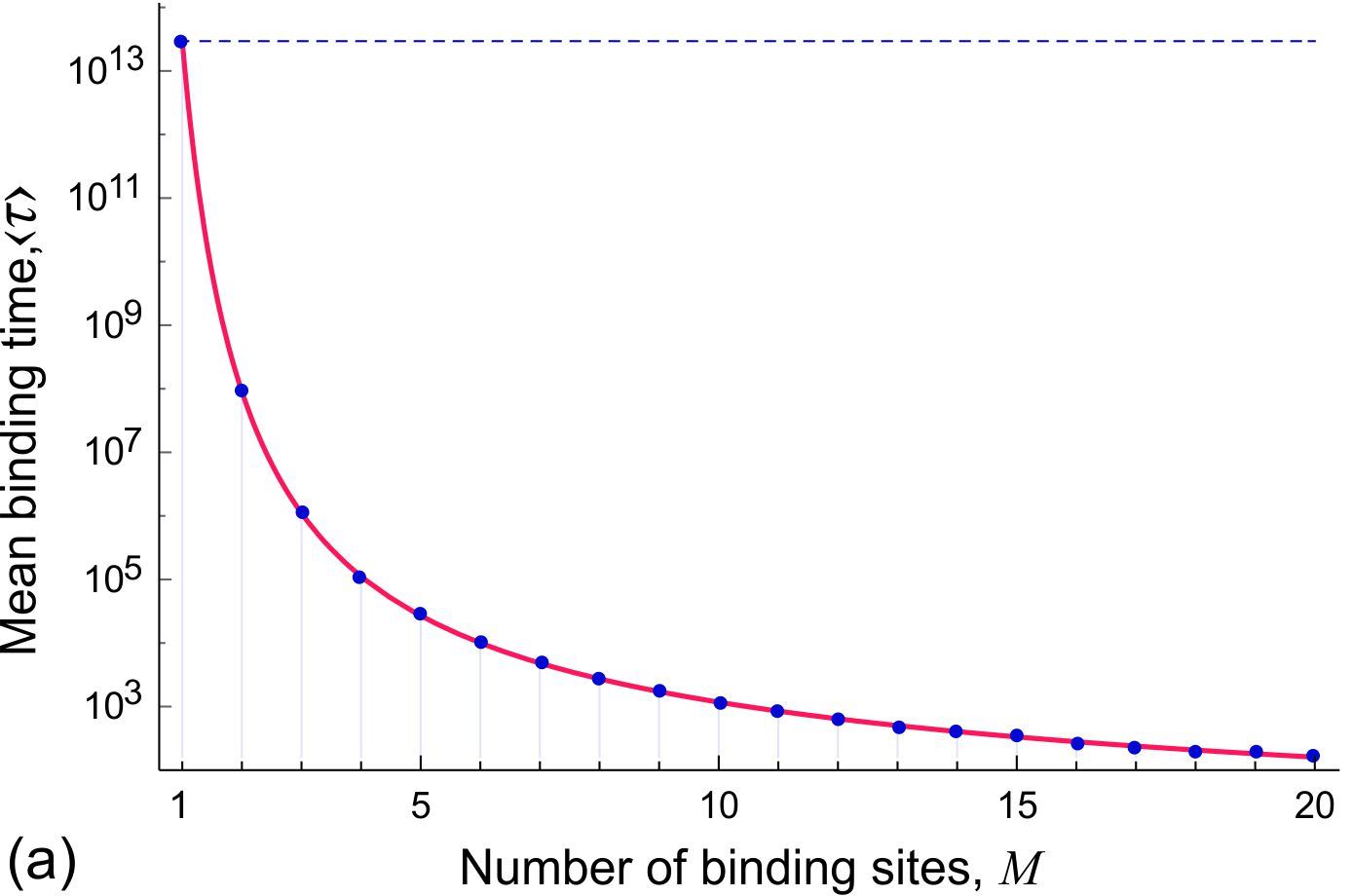}
   \includegraphics[width=0.45\textwidth]{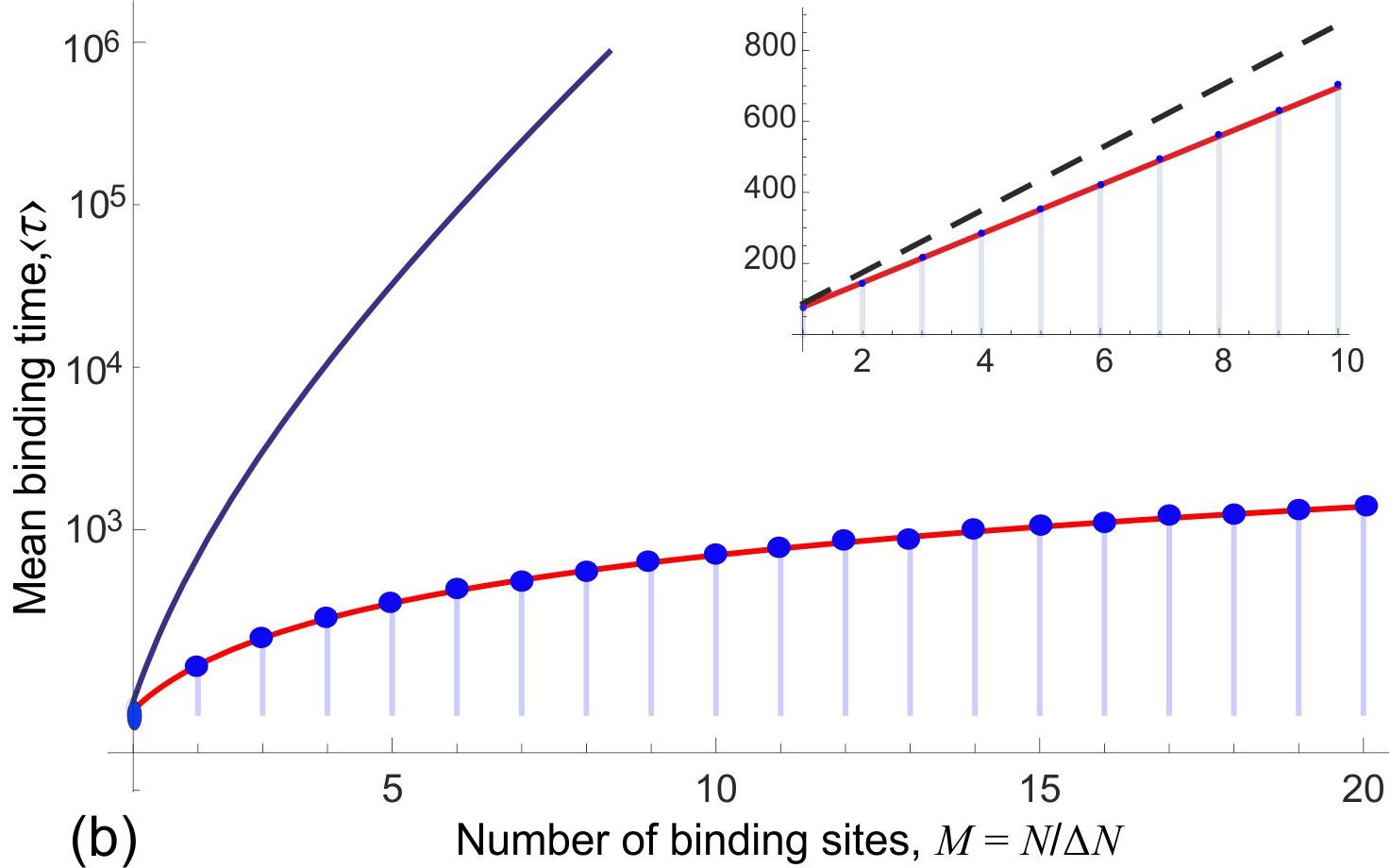}
    \caption{The mean time  (in units of $b^2/D$, logarithmic scale) to bind the chain to the surface, as a function of the number $M$ of  equidistant binding sites. (a) A fixed chain $N=100$ segments. The dashed blue line marks the case of $M=1$, when only the $N$th segment has a binding ligand, reaching for a receptor $a=40b$ away. As the number of binding sites along the chain increases, the time to bind the final receptor dramatically reduces. \ (b) A fixed interval between receptors  $\Delta N=20$, so $N=M \Delta N$. The plot compares a chain with single binding site at a distance $a=3 N b/20$ (solid blue line), and a chain with $M=N/\Delta N$ binding sites every 20 monomers, whose receptors are spaced at $\Delta a = a/M=3b$. In both cases, the end of the chain binds at the distance $a$. The inset illustrates that the typical binding time increases almost linearly with chain length or number of sites, in contrast to the exponential increase of this time for the single-site chain. The dashed black line indicates the line $M\tau_{1a}$, which is the strictly single-step zipper binding pathway. The possibility of occasional double steps lowers the binding time of an `accelerated zipper'.
In both plots, the blue dots represent the exact expression for $\langle \tau \rangle$; the continuous red line is the plot of \eqref{eq:tauapp}, where $\Delta a = a/M$.}
    \label{fig:sameN}
\end{figure}

This expression is the first main result of our paper. Figure \ref{fig:sameN} shows the comparison of the approximation presented in Eq. \eqref{eq:tauapp}, and the exact sum in Eq. \eqref{eq:taufull}, the latter plotted as discrete points at integer values of $M$. Evidently, the approximation (\ref{eq:tauapp}) is virtually indistinguishable from the exact average binding time, when the probability of making a double step is small.

How does adding more binding sites along a chain length influence its time to bind to a surface? Let us consider a chain of fixed length $N$, as usual grafted at the origin. There is a binding ligand on the end of the chain, and Eq. \eqref{eq:mfpt} gives the mean time for it to bind at a receptor on the surface a long distance $a$ away: $\tau_\mathrm{on}(N,a)$. Let us now add several more binding ligands on the chain, such that they have $N/M$ monomers in between, and the matching sequence of  equidistant receptor sites on the surface, such that they are a distance $\Delta a = a/M$ apart. 
 The resulting decrease in binding time is plotted in Fig. \ref{fig:sameN}(a). Note that the binding time is plotted on a logarithmic scale, so the effective decrease is quite dramatic when more binding sites are added to the chain. Equation \eqref{eq:tauapp} gives the scaling $\langle \tau \rangle \propto M^{-1}\exp [\alpha/M]$. 

We also examine the situation where the binding site density is kept constant, i.e binding sites on the chain are equally spaced, and the matching receptors on the surface are always spaced the same distance $\Delta a$ apart, but vary the total length of the chain. In this case the total chain length $N = M \Delta N$, and the distance to the last receptor is $a=M \Delta a$. The results are plotted in Fig. \ref{fig:sameN}(b) for the receptor density $\Delta a = 3\Delta N b/10 = 6b$. The comparison is made with the mean binding time for the chain with only one binding site at the end, with the chain length and the distance to the single receptor related in the same way: $a = 3 N b/10$ away, to illustrate the role of overall chain length. This time increases almost exponentially, see Eq. \eqref{eq:mfpt} giving  $ \tau_\mathrm{on} \propto N^{2}\exp [\alpha N]$. In contrast, the mean time to bind a sequence of receptors increases only $\sim$linearly with the chain length, illustrating that multiple sites massively enhance the binding rate. Note that a non-zero probability to make occasional double steps increases the binding rate even further, comparing with the straight zipper sequence, making it an `accelerated zipper' process -- this is illustrated by the linear plot in the inset of Fig. \ref{fig:sameN}(b).

\section*{Drift of center of mass}

\textcolor{black}{When we consider a single binding event, we do not consider the entire chain's length, but instead just the length between the tether and the binding site. However, in a sequence of binding events, we must consider how the rest of the chain moves around at the time of binding. In a typical monomer-by-monomer physisorption of a chain to a uniform surface, the rapid binding of monomers to the surface moves the effective current grafting point away from the center of mass of the remaining free chain, resulting in the  `stem and flower' configuration of Brochard-Wyart~\cite{Brochard1995,Bhattacharya2008}. This has an effect on the adsorption kinetics. Is there a similar effect when there is a section of chain between binding sites?}

\textcolor{black}{According to equilibrium polymer statistics, at the moment of a binding event, we would expect the remainder of the chain to be centered directly above the new grafting site. This stems from the Markovian treatment of the polymer chain, and is easily derived through Gaussian propagators. However, a study by Gu\'erin et al.~\cite{Voituriez2012} found that in reality, non-Markovian effects (i.e. non-Gaussian chain statistics before equilibrium is reached) play a large role in determining the dynamics of polymer configurations for reactions to a target in free space. The delay in reaching the equilibrium is often quite extended, so that the chain leaves the center of mass behind while reaching for a new target. This becomes clear if we consider the Rouse modes of the chain: the typical time for a monomer to fluctuate (travel a distance $a$) is much smaller than that for the chain center of mass to do the same. As such, the polymer chain is `left behind' when the next binding site finds its receptor, and we will need to consider the subsequent drift of the chain center of mass to the new grafting point. In this case, in contrast to the monomer-by-monomer physisorption, the rare binding steps result in a high stretching of the `stem' and a relatively high force pulling the remaining free chain towards the new equilibrium around the new grafting site. The mean time to diffuse a distance for the center of mass to diffuse the distance $a$ is:
\begin{equation}\label{eq:com}
\tau_\mathrm{com}=\frac{N_f a^2}{D},
\end{equation}
where $N_f$ is the number of monomers in the remaining free chain, so that $N_f\gamma$ is the effective friction constant for the free chain center of mass. As before, $D=k_BT/\gamma$ is the diffusion constant for a single monomer. In order for the Szabo-based~\cite{Szabo1980} expression for the binding time $\tau_{1a}$ to be valid, we should have a chain in equilibrium configuration, which occurs when $\tau_\mathrm{com}/\tau_{1a} \ll 1$.  This ratio takes the form:
\begin{equation}
\frac{\tau_\mathrm{com}}{\tau_{1a}}=9 \frac{N_f\varepsilon^2 \Delta a^2}{\Delta N^2b^4} e^{-3\Delta a^2/2\Delta N b^2}.
\end{equation} 
Assuming only a single-step zipper binding for simplicity, for the $m$th binding event, the remaining $N_f=(M-m+1)\Delta N$. Remembering that the total number of monomers, $N=M\Delta N$, and that the $M$th receptor is placed at a distance $a=M\Delta a$, this condition takes the form:
\begin{equation}\label{eq:m1}
m\gg M+1-\frac{MNb^4}{9\varepsilon^2 a^2} e^{3a^2/2MNb^2}.
\end{equation}
We can define the crossover point, $m_\mathrm{com}^*$, at which the binding time becomes comparable to the characteristic time of the center of mass diffusion by demanding equality in Eq. \eqref{eq:m1}.  So the equilibrium theory of binding to a distant site is valid at $m \gg m_\mathrm{com}^*$, and in the opposite limit the dynamics is determined by non-equilibrium (non-Markovian) statistics. }

How does this affect the zipper action? Let us assume that the polymer chain is initially equilibrated, with its center of mass close to the current point of grafting. After reaching for the next receptor site, the chain binds there, and then the remaining free chain finds its center of mass a distance $\Delta a$ out of equilibrium. The entropic force due to this stretching of the chain will provide an impetus to move the center of mass of the remaining chain to re-equilibrate \textcolor{black}{above the new grafting} position, and we can write down the dynamical equation for the movement of the center of mass:
\begin{equation}
-(N_f\gamma ) \dot{x}-\frac{3k_BT}{2N_fb^2}(x-\Delta a)=0,
\end{equation}
where, again, $N_f$ is the number of monomers in the remaining free chain. It follows that the relaxation time to the new equilibrium of the free chain is given by
\begin{equation}\label{eq:relT}
\tau_\mathrm{drift}=\frac{2 N_f^2 b^2}{3 D}.
\end{equation}

Then, foillowing a similar method to before, we can find a condition on $m$ for the drift time to be dominant, $\tau_{\mathrm{drift}}/\tau_{1a}\gg 1$:
\begin{equation}\label{eq:m}
m\ll M+1-\frac{1}{\sqrt{6}}\frac{b}{\varepsilon}e^{3a^2/4MNb^2},
\end{equation}
because for early binding events (small $m$), there is a lot of free chain, and so $\tau_\mathrm{drift}$ is large. For later binding events, the remaining free chain is able to equilibrate fast, and so there is no need to account for the drift of center of remaining mass.

\textcolor{black}{We can define the other crossover point, $m_\mathrm{drift}^*$, at which the chain binding changes from being limited by the chain relaxation time, given by Eq. \eqref{eq:relT}, to being limited by the time $\tau_{1a}$ to reach the next binding site, by setting equality in \eqref{eq:m}.  It turns out that, in spite of subtle differences, the crossover expressions  are quite close numerically: $m_\mathrm{com}^*\approx m_\mathrm{drift}^*=m^*$.  So when the equilibrium expression for the binding time is valid ($m \gg m_\mathrm{com}^*$) it is also the case that the total binding time is dominated by the reaching time. On the other hand, when the chain is not equilibrating fast enough ($m \ll m_\mathrm{com}^*$), it is also the case that the binding is limited by the slow drift of the chain center of mass.}

Therefore, the first $m^*$ binding events (when the free chain segment is still long) will be relaxation-limited, while the last $(M-m^*)$ events are independent of the chain length. The effective  binding time takes the form:
\begin{equation} \label{eq:adT}
\tau = \sum_{m=1}^{m^*}\tau_\mathrm{drift}(m)+(M-m^*)\tau_\mathrm{1a}
\end{equation}
(neglecting the weaker effect of accelerated zipper, for clarity). 
If $m^*<1$, which is always the case at small $M$, then all binding events are reach-limited, and the expression returns to the simple linear zipper $\tau=M\tau_\mathrm{1a}$.

\begin{figure}
	\centering
	\includegraphics[width=0.45\textwidth]{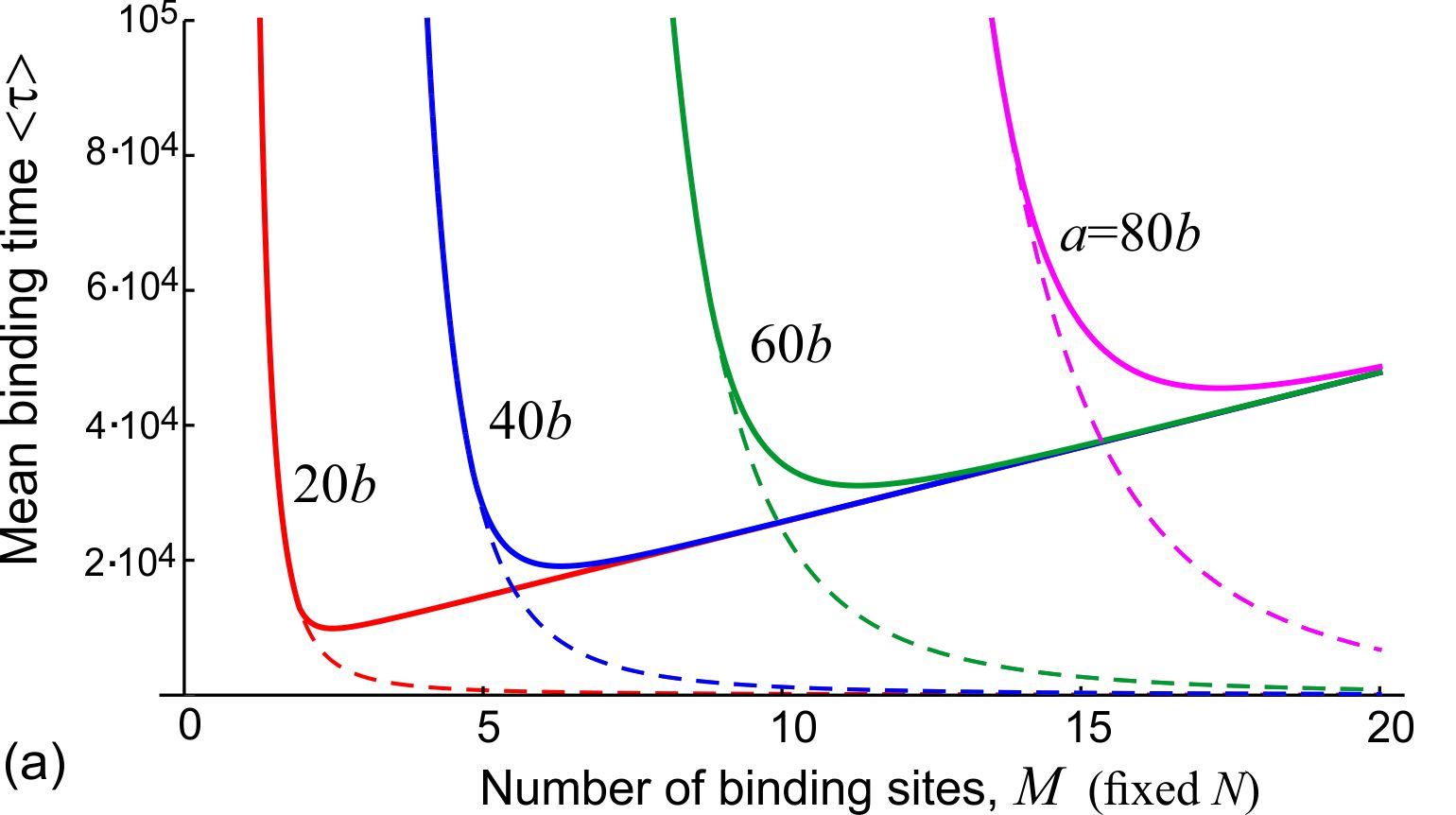}
	\includegraphics[width=0.45\textwidth]{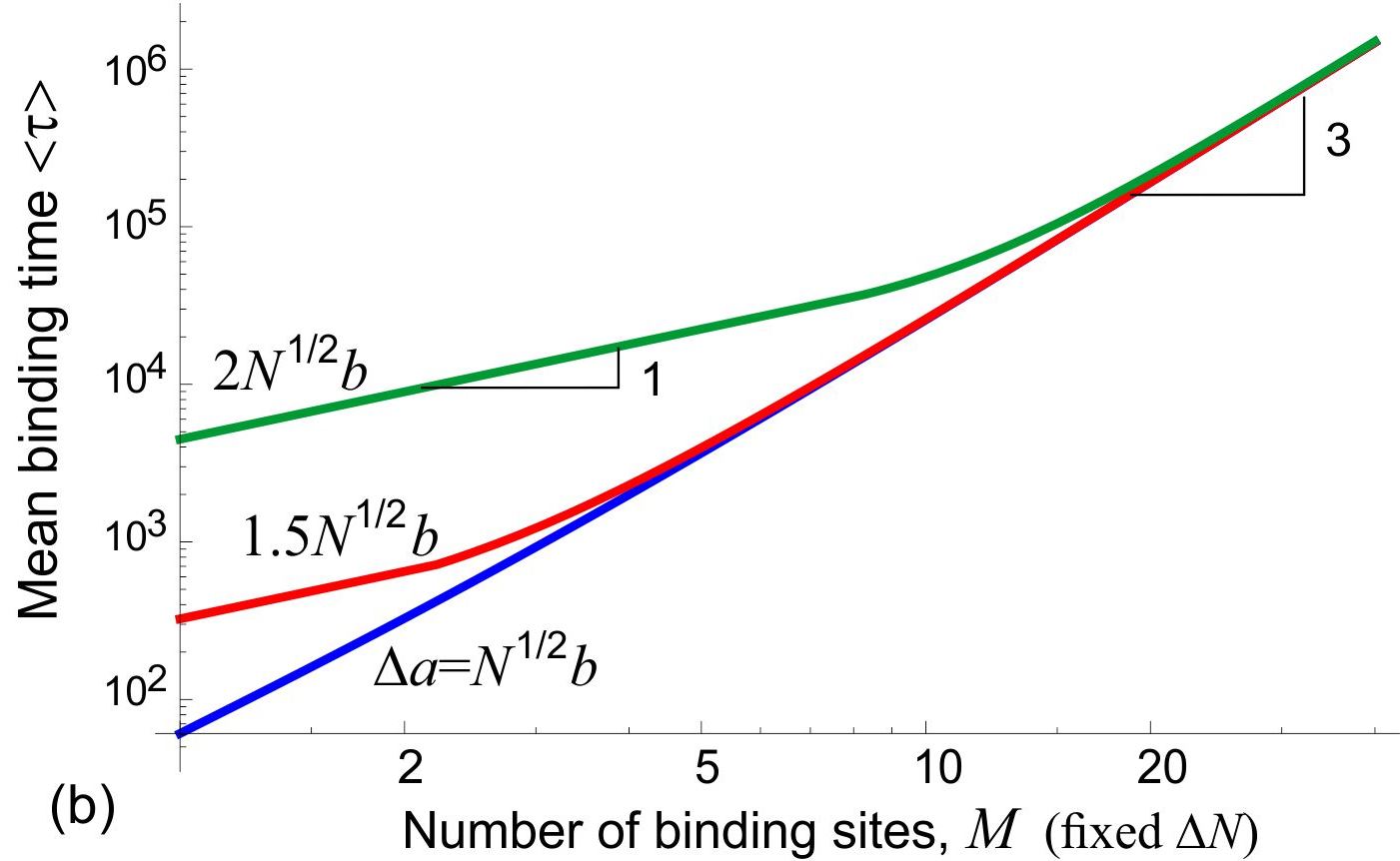}
	\caption{The adhesion time is plotted in units of $b^2/D$, for $\varepsilon=b$. (a) The linear plot for a fixed $N=100$. The set-up is $M$ receptors, with the final receptor placed at varying distances $a$ away in different curves (labelled on the plot). The dashed line shows how the reaction-limited time reduces as $M$ increases. (b) The log-log plot, for fixed $\Delta N=N/M=10$. From top to bottom, the distance between consecutive receptors $\Delta a=\sqrt{\Delta Nb^2}$, $1.5\sqrt{\Delta Nb^2}$, and $2\sqrt{\Delta Nb^2}$.  For shorter chains, there is a reach-limited linear `zipper' region (gradient slope of 1 is shown), before switching to the cubic increase of time with length as chains get longer (gradient slope of 3 is shown), see Eq. \eqref{eq:cubic}. For larger $\Delta a$, the linear region is extended.}
	\label{fig:Mdiff}
\end{figure}

Figure~\ref{fig:Mdiff}(a) illustrates how increasing the number of intermediary binding sites affects the total time to adsorb a chain, which is essentially the time to bind to the final receptor a fixed distance away. This is analogous to Fig.~\ref{fig:sameN}(a) obtained in the fully equilibrium-chain setting; in fact, the dashed lines in Fig. \ref{fig:Mdiff}(a) give the lines of Eq. \eqref{eq:tauapp} as in  Fig.~\ref{fig:sameN}(a). We see that for small $M$, increasing the number of binding sites along the chain reduces the total binding time, because the process is purely reach-limited ($m^*<0$). As the number of intermediary sites increases, however, we cross into a regime limited by the relaxation (drift) of the chain center of mass. Here, increasing the number of sites lowers the drift force in the Langevin equation, and so the chain will actually take longer to reach the terminal receptor.  Note that all curves saturate on the same line, because the relaxation-limited time does not depend on $a$: making the summation in Eq. \eqref{eq:adT} for $m^*=M$ gives a linear estimate $\tau \approx M(2Nb^2/9D)$ for this section of the curves in Fig. ~\ref{fig:Mdiff}(a).

Between these two effects: the zipper time, which becomes shorter when more receptors are added on a fixed interval, and the relaxation (drift) time of the chain center of mass, there is clearly an optimal number of binding sites that achieves the shortest time for the complete chain adhesion. To find this point, one has to solve the derivative of Eq. \eqref{eq:adT}: $d \tau(M)/dM =0$. This is a complicated algebraic task, which simplifies in the limit  $a^2>Nb^2$, that is, when the chain needs to absorb by stretching over some distance. In this case the minimum of the adhesion time is for the $M$ determined by the equation:
\begin{equation}
16N\varepsilon^3M^4-2\sqrt{6}a^2be^{\frac{9a^2}{4MNb^2}}=0. 
\end{equation}
This transcendent equation has a solution in terms of the Lambert W-function, or the product logarithm, but this is too cumbersome to get a clear understanding. However, in the limit  $a^2 \gg Nb^2$, the leading contribution to the `optimal' number of receptors for the shortest adhesion time takes the form
\begin{align}
M_\mathrm{opt} \approx & \frac{3a^2}{4Nb^2}\left[\ln\left(\frac{9a^2}{16Nb^2}\frac{\varepsilon}{b}\sqrt{\frac{3}{2}}\right)\right]^{-1} \nonumber \\
= & \frac{3a^2/4R_g^2}{  \ln\left(0.69 [a^2/R_g^2] (\varepsilon/b) \right)} .
\end{align}

In Figure~\ref{fig:Mdiff}(b) we instead fix the distance between receptors, $\Delta a$, and the chain length between binding sites, $\Delta N$, as in Fig.~\ref{fig:sameN}(b). When we plot the total adhesion time on a log-log scale, one can clearly see the two distinct regimes: the reach-limited at small $M$ (the linear zipper increase), changing to the cubic increase as we go to larger $M$ and the relaxation-limited regime. In this latter regime, Eq. \eqref{eq:relT} with $N_f=[M-m+1]\Delta N$ dominates the contributions to the total adhesion time \eqref{eq:adT}, producing the dominant cubic dependence on the number of receptors, as illustrated in the plot:
\begin{equation} \label{eq:cubic}
\tau \approx \sum_{m=1}^{M}  \frac{2 (N-[m-1]\Delta N)^2 b^2}{3 D} \  \to \   \frac{2 M^3 \Delta N^2 b^2}{9 D} +...
\end{equation}

\section*{Conclusions}

Experiments will usually determine the rate constant of a reaction, given by the reciprocal of the binding time. For the single binding event of a tethered ligand: 
\begin{equation}\label{eq:kfin}
k_\mathrm{on}=\frac{1}{\tau_\mathrm{on}}=\frac{9D\varepsilon^2}{N^2b^4}e^{-3a^2/2Nb^2}.
\end{equation}
The rate constant defined by Eq. (\ref{eq:kfin}) is Arrhenius in form ($k_\mathrm{on}=Ae^{-F/k_{B}T}$) if we take the free energy barrier to be purely entropic: $F=-T\Delta S$, where
\begin{equation}
\Delta S = -\frac{3k_{B}}{2Nb^2}a^2
\end{equation}
is the reduction in entropy for an entropic spring of length $Nb$  stretched from $0$ to $a$. We might then, naively, believe that the binding time will increase monotonically as the length of chain increases -- the entropic penalty will become smaller and smaller.

However, as we find in Eq. (\ref{eq:mfpt}), there is another competing effect that increases the mean first passage time: as the chain gets longer, the effective volume that the site can explore relative to the receptor volume also increases. If we increase the chain to an infinite length, we actually return to a free particle scenario, and there is not enough confinement for the end of the chain to ever hit the receptor. For two binding sites placed a distance $a$ apart, we can easily find the optimal length of tether chain for the fastest time to bind:
\begin{equation}
N^* = \frac{3a^2}{4b^2} , \qquad \mathrm{hence} \ \ \ \tau_\mathrm{on}^*= 0.46 \frac{a^4}{D\varepsilon^2} .
\end{equation}

The difference in scaling of the mean binding time with receptor size $\varepsilon$ in the tethered chain is an interesting feature, especially when compared with the looping chain or `narrow escape' scenarios (which both has the $\varepsilon^{-1}$ scaling). When we move from an unrestricted free space to a half-space, the polymer exclusion boundary condition effectively reduces the ways in which the chain can approach the receptor. This effect is purely entropic -- it is the reduction in possible ways that the chain can orient itself near the wall that `prevents' the chain from touching the wall.  Together with the entropic barrier, this effect determines the binding time of a tethered ligand. 

Here we approached the `reaching for the target' problem in a way completely different from the `narrow escape' problem of Holcman et al. Instead of examining the Smoluchowski problem in an effective potential imposed by the constraints, we have generated the mean-field equilibrium probability density $P_\mathrm{eq}(\rho)$, and were able to utilize the mean first passage time solution of Szabo et al. It may well be that this approach generates analytical / numerical solutions more rapidly even for the more complex problems invoilving potential interaction and non-ideal polymer chain, as well as confined Brownian particles, since we are not having to look at the dynamical effects -- only at how these affect the equilibrium effective potential.

In the problem of chain adhesion to a sequence of binding sites on a surface, we find that the addition of intermediary binding sites along a chain length has a dramatic `zipper effect', massively decreasing the time for the chain to bind fully along its length. When we examine our main result -- the expression for the average binding time in Eq. \eqref{eq:tauapp}, it is important that for large $\Delta a$ and large $M$, the dominant binding process is the single-step `zipper' pathway with the mean time approximately equal to $M\tau_{1a}$, as we can see in the inset of Fig.~\ref{fig:sameN}(b). Only occasionally will a chain bind with a double step, and this correction to the binding time also scales linearly with $M$, see Fig. \ref{fig:sameN}(b). It is not until receptors are very tightly grouped (small $\Delta a$) that double-step processes start to become relevant. This is suggestive of reality -- if a polymer chain has a specific substrate structure to bind to, then steric effects may well force the polymer to bind in a very conserved and controlled sequence (as in nature), just by virtue of the high entropic penalty for binding `out of order'.

The existence of an optimal number of intermediate receptors, $M_\mathrm{opt}$ for the shortest time of full chain adhesion is our second main result. This is an interesting feature, perhaps contrary to an expectation that by reducing the reaction barrier (and the associated individual binding time) one would increase the overall rate of adsorption. For chains with many intermediate receptors, although there is fast attachment to each individual site, the process of moving the center of mass of the remaining free chain down the line of receptors is slow, because the entropic pulling force causing this drift is weak. Conversely, if there are only a few receptors, even though any attachment event will provide a strong force to move the free chain to its new equilibrium position, the binding time itself is prohibitively slow.

It would be interesting to apply even this basic solution to a practical problem of amyloid assembly, where the new peptide subunit has to bind to a specific sequence of sites by hydrogen-bonding the $\beta$-sheet at the end of the existing filament \cite{Buell2012}, and the entropic barriers are explicitly reported.

\subsection*{Acknowledgments}
{S.B. appreciates several helpful discussions with Preeyan Parmar and John Grenfell-Shaw. This work was supported by EPSRC.}

%\bibliographystyle{apsrev4-1}
%\bibliography{zipbib}

\newpage

\begin{appendix}
\section{Calculation of radial distribution functions}  \label{app:A}

To get an effective propagator in one dimension, we first transform \eqref{eq:propend} into spherical coordinates:

\begin{equation}
x - a = \rho\sin\theta\cos\phi\,, \;  
y = \rho\sin\theta\sin\phi\, , \; z = \rho\cos\theta.
\end{equation}
The two scalar products in the combined exponents of Eq.(\ref{eq:propend}) become:
\begin{equation}
(\mathbf{r}\pm b\hat{\mathbf{z}})^2=a^2+b^2+\rho^2+2a\rho\sin\theta\cos\phi\pm2b\rho\cos\theta .
\end{equation}

The next step of integration over the solid angle on the unit hemisphere is not easy. We need to evaluate
\begin{equation}\label{eq:intI}
I = \int_{0}^{\pi/2}d\theta\sin\theta\int_{0}^{2\pi}d\phi \;e^{-\alpha\cos\phi\sin\theta\pm\beta\cos\theta}, 
\end{equation}
where parameters $\alpha$ and $\beta$ involve $N,b,a$ and $\rho$.
This is solved by realizing that the integrand has a non-trivial axial symmetry. We transform back into Cartesian coordinates about the target: $x'=\cos\phi \sin\theta$, $z'=\cos\theta$, and rotate these new coordinates by an angle $\varphi=\tan^{-1}(\beta/\alpha)=\tan^{-1}(b/a)$ around the $y$-axis. The direction of this rotation depends on the sign of the $z$-term in the exponent (i.e. whether we are dealing with the `real' or `image' Gaussian term). This rotation means we are essential integrating $\exp(-\sqrt{a^2+b^2}x)$. However, we must be careful when we define the surface we are integrating over. Since we aren't integrating over a line in the plane of the surface, the hemisphere appears tilted with respect to the variable of integration $x$, as in Figure. The surface element is given by $\psi(x)\sqrt{1-x^2}ds$. On the unit sphere $y=\pm\sqrt{1-x^2}$, and the element $ds$ is given by
\begin{equation}
ds = \sqrt{1+\left(\frac{\partial y}{\partial x}\right)^2}dx = \frac{dx}{\sqrt{1-x^2}}.
\end{equation}
The opening angle of the surface element has the exact expression 
\begin{equation}
\psi(x)=
\begin{cases}
2\pi, & -1\leq x<-\cos\varphi \\
\pi-2\sin^{-1}\left(\frac{x\tan\varphi}{\sqrt{1-x^2}}\right), & -\cos\varphi\leq x\leq cos\varphi \\
0, &cos\varphi<x\leq1
\end{cases}
\end{equation}
The radial probability density is defined according to the normalisation over the allowed half-space:
\begin{equation}
\int_{0}^{\infty}d\rho\; 2\pi \rho^2 P_{eq}(\rho)=1
\end{equation}
where $2\pi \rho^2$ accounts for the area of a hemispherical shell of radius $\rho$. Using this, we find that the radial probability density takes the form
\begin{align}\label{eq:pradfull}
&P_{eq}(\rho)=\sqrt{\frac{N\pi}{6}}\left(\frac{3}{2\pi Nb^2}\right)^{3/2}e^{-\frac{3(a^2+b^2+\rho^2}{2Nb^2}}\times \nonumber \\
&\left[\frac{2 Nb^2}{3\rho\sqrt{a^2+b^2}}\left[\cosh\left(\frac{3\sqrt{a^2+b^2}}{2Nb^2}\rho\right)-\cosh\left(\frac{3\rho a}{2Nb^2}\right)\right]\right. \nonumber \\
&+\left.\frac{2 b}{a}I_{1}\left(\frac{3\sqrt{a^2+b^2}}{Nb^2}\rho\right)-\frac{4b^2}{\pi a^2}\sinh\left(\frac{3\sqrt{a^2+b^2}}{Nb^2}\rho\right)\right].
\end{align}

\section{Calculation of looping time}  \label{app:B}
Here, we show that the mean looping time of a polymer, as calculated from our method, coincides with the expression derived by Szabo et al~\cite{Szabo1980}. If we consider first the more general problem of a chain with one end `tethered' in place (in reality, since we can change our frame of reference of the polymer, we may do this for the free chain). Then, the distribution of the `free end' is given by
\begin{equation}
P(\mathbf{r})=\left(\frac{3}{2\pi Nb^2}\right)^{3/2}\exp\left(-\frac{3\mathbf{r}^2}{2Nb^2}\right)
\end{equation}
We want to calculate the time for the free end to hit a sphere of radius $\varepsilon$ centered a distance $a$ from the first monomer. As in Appendix A, we may transform into spherical polar coordinates and then integrate over the polar angles to obtain a probability distribution over $r$. This integral is very similar in form to \eqref{eq:intI}, but with the integration limits extended over the entire unit sphere, and $\beta=0$
\begin{equation}
I = \int_{0}^{\pi}d\theta\sin\theta\int_{0}^{2\pi}d\phi \;e^{-\alpha\cos\phi\sin\theta}
\end{equation}
Tranforming back into Cartesian coordinates, we have \begin{equation}
I = 2\pi\int_{-1}^{1}dx\; e^{-\alpha x}=\frac{2\pi}{\alpha}\left(e^\alpha-e^{-\alpha}\right).
\end{equation}
As such, the resulting radial probability distribution about the target is 
\begin{equation}
P_{eq}(\rho)=\frac{1}{2\pi a\rho}\sqrt{\frac{3}{2\pi Nb^2}}\left(e^{-\frac{3(\rho-a)^2}{2Nb^2}}-e^{-\frac{3(\rho+a)^2}{2Nb^2}}\right).
\end{equation}
When we use this probability distribution in the expression for mean first passage time
\begin{equation}
\tau = 2\pi \int_{\varepsilon}^{\infty}d\rho\left[D \rho^2 P_{eq}(\rho)\right]^{-1}\left[\int_{\rho}^{\infty}d\rho'\; \rho'^2 P_{eq}(\rho')\right]^2 , 
\end{equation}
we find that the integral, though not analytically solvable, is dominated by the value of the integrand at small $\rho$. As $\rho\to 0$, the probability distribution tends to a non-zero constant, and so we can make the approximation
\begin{align}
\tau_\mathrm{on} &\approx \sqrt{\frac{\pi}{54}} \frac{(Nb^2)^{3/2}}{D}e^{\frac{3a^2}{2Nb^2}}\int_{\varepsilon}^{\infty}\frac{d\rho}{\rho^2} =
 \sqrt{\frac{\pi}{54}} \frac{(Nb^2)^{3/2}}{D\varepsilon}e^{\frac{3a^2}{2R_g^2}}.
\end{align}
From here, it is a matter of setting $a=0$ to recover the Szabo result for the looping time of a polymer in three dimensions, shown in \eqref{eq:sza}.

\section{Calculation of multiple-binding time}  \label{app:C}

Evaluating the sums in \eqref{eq:tave} requires that we calculate the average $ \langle n \rangle$:
\begin{equation}\label{eq:tau15}
\langle\tau\rangle = \frac{M - \langle n \rangle_{p_n}}{k_1+k_2} .
\end{equation}
The average $\langle n \rangle = \sum n P(n)$ calculates directly to produce an exact expression
\begin{equation}
\langle n \rangle = \frac{p_2\left((M-1)p_2^{M+1}+(M+1) p_2^M+(M+1)p_2+M-1\right)}{(p_2+1)^2 \left(p_2^{M+1}+1\right)}, \nonumber
\end{equation}
where $p_2 = 1-p_1$ is the probability of making a double step given in \eqref{eq:p12}. When the probability of making a double step is small, $p_2\ll 1$, then we can expand the average number of double steps $\langle n\rangle$ in a Taylor series in powers of $p_2$, prodicing 
\begin{align}
\langle n \rangle &= (M-1)p_2 - (M-3)p_2^2+(M-5)p_2^3+... \nonumber \\
&=\sum_{n=1}^{M/2}(-1)^{n-1}[M-(2n-1)]p_2^{n}
\end{align}

We arrive at the approximate expression given in \eqref{eq:tauapp} by considering the leading order term in the sum, and substituting explicit expressions for $p_2$, $k_1$ and $k_2$:
\begin{align}
\langle \tau\rangle &\approx \frac{p_2 +M(1-p_2)}{k_1+k_2} \nonumber \\
&=\frac{1+4 M e^{\frac{3 \Delta a^2 M}{2 b^2 N}}}{\left(1+4 e^{\frac{3 \Delta a^2 M}{2 b^2 N}}\right)^2}\cdot \frac{4b^4N^2}{9D \varepsilon^2 M^2}e^{\frac{3\Delta a^2M}{Nb^2}}
\end{align}
which is just \eqref{eq:tauapp}.
\end{appendix}

\end{document}